\documentclass[showpacs,aps,amssymb,floatfix,prd,preprintnumbers]{revtex4}
\usepackage{epstopdf}
\usepackage{capt-of}
\usepackage{graphicx}  
\usepackage{dcolumn}   
\usepackage{bm}
\usepackage{amsmath}
\usepackage[font=scriptsize]{caption}
\usepackage[colorlinks]{hyperref}
\begin{document}
\title{\bf Bouncing cosmology in extended gravity and its reconstruction as dark energy model}

\author{A. S. Agrawal \footnote{Department of Mathematics,
Birla Institute of Technology and Science-Pilani, Hyderabad Campus,
Hyderabad-500078, India, E-mail: agrawalamar61@gmail.com}, Francisco Tello-Ortiz \footnote{Departamento de F\'isica, Facultad de Ciencias Básicas, Universidad de Antofagasta, Casilla 170, Antofagasta, Chile, Email:francisco.tello@ua.cl}, B. Mishra \footnote{Department of Mathematics,
Birla Institute of Technology and Science-Pilani, Hyderabad Campus,
Hyderabad-500078, India, email: bivu@hyderabad.bits-pilani.ac.in}, S.K. Tripathy \footnote{Department of Physics, Indira Gandhi Institute of Technology, Sarang, Dhenkanal, Odisha-759146, India, tripathy\_sunil@rediffmail.com}}

\affiliation{ }

\begin{abstract}
\textbf{Abstract:}
In this paper, we have presented a bouncing cosmological model of the Universe in an extended theory of gravity. The dynamical behaviour of the model obtained from the flat FLRW space-time along with the violation of null energy condition have been shown. The geometrical parameters show singularity behaviour at the bouncing epoch. The parameters involved in the scale factor play a major role in the bouncing behaviour. In addition, the coupling parameter that resulted in the minimal matter-geometry coupling in the extended gravity has significant role to avoid the singularity of equation of state parameter at the bouncing epoch. Using a linear homogeneous perturbation calculation, we show the stability of the model. 
\end{abstract}

\maketitle
\section{Introduction}

The standard cosmological model suffers from issues like initial singularity, the horizon problem, flatness problem and baryon asymmetry, though it has been successful in explaining the early Universe. The inflationary theory has been successful in solving some of these issues of  early Universe, but it still suffers from the singularity problem and the trans-Planckian problem in terms of fluctuation. According to the inflationary scenario, the Universe experienced an exponential expansion at an initial epoch. As usual the singularity occurs before the onset of inflation and therefore the inflationary scenario fails to recreate the complete past history of the universe. As a possible solution to the inflationary scenario problem, the matter bounce scenario has been suggested \cite{Bars11,Brandenberger12}. According to the matter bounce scenario, it is believed that the Universe had a contracting phase and is capable of expanding without encountering any initial singularity. That is, the Universe undergoes an initial matter-dominated contraction phase followed by a non-singular bounce and then there is a causal generation for fluctuation. The cosmological models replace big bang cosmological singularity  with a big bounce scenario, which is a smooth transition from contraction to expansion phase \cite{Brandenberger11, Elizalde15}. But, one important observation is that in a flat Universe, the presence of non-singular bounce may lead to the violation of null energy condition. In fact the violation of null energy condition can be experienced  in generalised Galileon theories that supports the possibility on non-singular cosmology \cite{Kobayashi16}. Another issue in the bouncing models is that in many occasions the models are seen to be unstable. However, it has been argued that, beyond Horndeski theory and effective field theory, the stable bouncing cosmologies may be framed \cite{Creminelli16, Kolevatov17, Cai17}. The occurrence of big bounce scenario to replace the big bang singularity might be an interesting topic in the context of modified theories of gravity. Several bouncing cosmological models are presented in recent times in modified theories of gravity,  such as $f(R)$ gravity, \cite{Barragan09,Saidov10,Bamba14,Amani16,Das18,Chakraborty18}, $f(R,T)$ gravity \cite{Shabani18, Singh2018, Tripathy2019, Mishra19,Tripathy21}, teleparallel gravity \cite{Cai11,Logbo19}, $f(Q,T)$ gravity \cite{Agrawal21}, $f(T,B)$ gravity \cite{Caruana20} modified Gauss-Bonnet gravity \cite{Oikonomou15,Bamba15,Shamir21} and so on. Besides the proposal of matter bounce scenario in different modified theories of gravity, there have been many proposals for a smoothening of the slow contraction process which may include both the classical as well as quantum mechanical treatments \cite{Cook2020, Albaran2017, Ijjas19, Ijjas20}.\\

In order to include the dark energy era and to address the late time cosmic acceleration issue in the matter bounce scenario, the loop quantum cosmology approach may be adhered to \cite{Bojowald09,Ashtekar11,Cai11,Cailleteau12,Quintin14}. The matter bounce scenario can be deformed to the occurrence of late time acceleration, so it can also be termed as the deformed matter bounce scenario. In the deformed matter bounce scenario, at $t\rightarrow\infty$, the contracting phase of Universe starts and then it bounces off at $t=0$ and subsequently expands again. Only at late time, the cosmological evolution gets affected by the deformation of matter bounce Universe. The motivation of the singular deformation of matter bounce basically from the fact that the Universe undergoes a late time acceleration era and the equation of state parameter $\omega_{eff}\approx-1$ and this feature is not there in the standard matter bounce scenario. Only at late time, the standard matter bounce scenario and deformed matter bounce scenario behave differently \cite{Odintosv16}. Another aspects of the matter bounce research is its consistency with the observations. When the potential function is included in the action, the scalar field remains on a flat plateau of the potential function after the bounce and hence the Universe experiences inflation. Interestingly, during the slow-roll inflation, the flat potential is required for experiencing the bounce. The bouncing solution becomes consistent if the scalar field continues to slow-roll for a sufficiently long time and finally finds another minimum with a sufficiently small and positive cosmological constant \cite{Matsui19}.\\

In this paper, the motivation is to frame the bouncing cosmological model of the Universe in an extended theory of gravity with a suitable bouncing scale factor. Given a bouncing scenario, we have studied the cosmic dynamics of the model. The paper is organised as follows: In Sec-II, the basic formalism and the field equations of $f(R,T)$ gravity have been presented. The bouncing model, the parameters such as dynamical parameters, energy conditions, cosmographic parameters, stability analysis are given in Sec-III. In Sec-IV, to connect with the present observational value of the Hubble parameter, we have reconstructed the model by the redshift parameter and obtained the value of the free parameters. The conclusion of the work has been given in Sec-V.

\section{$f(R,T)$ Gravity Field Equations in FRW metric} 
The action of $f(R,T)$ gravity, where $R$ and $T$ be respectively the Ricci scalar and trace of energy momentum tensor ($T_{ij}$) takes the form,
\begin{equation}\label{eq.1}
S=\int{\left[\frac{f(R,T)}{16\pi}+\mathcal{L}_{m} \right]}\sqrt{-g}d^{4}x,
\end{equation}
where $\mathcal{L}_{m}$ be the matter Lagrangian and we define the stress-energy tensor of matter as,
\begin{equation}\label{eq.2}
T_{ij}=-\frac{2}{\sqrt{-g}}\frac{\delta(\sqrt{-g}\mathcal{L}_{m})}{\delta g^{ij}}.
\end{equation}
We have considered here the non-minimal matter geometry coupling as, $f(R,T)=f_{1}(R)+f_{2}(T)$. Varying action \eqref{eq.1} with respect to the metric tensor $g_{ij}$, the field equations of $f(R,T)$ gravity with non-minimal matter coupling can be obtained as \cite{Harko11},
\begin{equation}
f_{R}(R)R_{ij}-\frac{1}{2}f(R)g_{ij}-(\nabla_{i}\nabla_{j}-g_{ij})f_{R}(R)=8\pi T_{ij}+f_{T}(T)T_{ij}+[f_{T}(T)p+\frac{1}{2}f(T)]g_{ij}. \label{eq.3}
\end{equation}

In eqn. \eqref{eq.3} we denote, $f_{R}(R)=\partial f_{1}(R)/\partial R$ and $f_{T}(T)=\partial f_{2}(T)/\partial T$ and $p$ be the pressure of the matter. Among the three choices of $f(R,T)$ proposed (Harko et al. \cite{Harko11}); we have considered, $f(R,T)=R+2f(T)$. There are good number of cosmological models presented in literature with $f_1(R)=R$ and $f_2(T)=\beta T$, $\beta$ being the coupling constant \cite{Shamir15,Moraes15,Tripathy20}. Here we wish to incorporate the time independent cosmological constant $\Lambda_{0}$ in $f_2(T)$, such that $f(R,T)=R+2\beta T+2\Lambda_0$ \cite{Mishra18,Tarai20}. We consider a flat FLRW space time
\begin{equation}\label{eq.4}
ds^{2}=dt^{2}-a^{2}(t)[dx^2+dy^2+dz^2],
\end{equation}
where $a$ is the scale factor of the universe. The field equation \eqref{eq.3} can now be expressed as,
\begin{equation}\label{eq.5}
G_{ij}=(8\pi+2\beta)T_{ij}+\Lambda(T)g_{ij},
\end{equation}
where $\Lambda(T)=(2p+T)\beta+\Lambda_{0}$ be the effective time variable cosmological constant. It is to mention here that post supernovae observation, the role of cosmological constant $\Lambda$ has become important in the study of accelerating cosmological model; prior to this $\Lambda$ was assumed to be zero. However, in the present extended gravity theory, it varies with the evolution of Universe and therefore comes out as a function of cosmic time. Interestingly, $\Lambda$ reduces to a pure constant $\Lambda_0$ for a vanishing $\beta$. Now the field equations \eqref{eq.5} can be reduced to,

\begin{equation}\label{eq.6}
G_{ij}=(8\pi+2\beta)T_{ij}+[(2p+T)\beta+\Lambda_{0}]g_{ij}.
\end{equation}

We assume a non-dissipative perfect fluid distribution in the Universe for which the energy momentum tensor is expressed as,
\begin{equation}\label{eq.7}
T_{ij}=(p+\rho)u_{i}u_{j}-pg_{ij}.
\end{equation}
Consequently the field equations of $f(R,T)$ gravity  \eqref{eq.6} can be obtained as,
\begin{eqnarray}
2\dot{H}+3H^{2}&=& - \eta p+\beta \rho +\Lambda_{0},   \label{eq.8}  \\
3H^{2}&=&\eta \rho- \beta p+\Lambda_{0}.    \label{eq.9}
\end{eqnarray}
where $\eta=8\pi +3\beta$ and $H=\frac{\dot{a}}{a}$ is the Hubble parameter. In the above equations, an over dot represents an ordinary derivative with respect to cosmic time. Performing some algebraic manipulations among eqns. \eqref{eq.8} and \eqref{eq.9}, we can derive the pressure $p$, energy density $\rho$ and equation of state (EoS) parameter $\omega=\frac{p}{\rho}$ in term of Hubble parameter as,

\begin{eqnarray}
p&=& -\frac{1}{(\eta^2-\beta^2)}\left[2\eta\dot{H}+3(\eta-\beta)H^2-(\eta-\beta)\Lambda_0\right], \label{eq.10} \\ 
\rho&=& \frac{1}{(\eta^2-\beta^2)}\left[-2\beta \dot{H}+3(\eta-\beta)H^2-(\eta-\beta)\Lambda_0 \right], \label{eq.11}\\ 
\omega&=&-1+\left[\frac{2(\eta+\beta) \dot{H}}{2\beta \dot{H}-3(\eta-\beta)H^2+(\eta-\beta)\Lambda_0}\right].\label{eq.12}
\end{eqnarray}

The motivation behind expressing the parameters in Hubble term is that we wish to study the bouncing behaviour of the model, with a widely known bouncing scale factor $a$. In classical bounce, the Universe bounces after being reduced to a small finite size. The energy density should appear below the Planck scale to neglect the effect of quantum gravity. This type of transition can occur with the violation of null energy condition (NEC) in finite period of time including the bouncing epoch. So, in the subsequent section we shall present the model with the bouncing scale factor. 

\section{The Bouncing Model and The Analysis}\label{sect-iii}

The violation of NEC prescribed that with increase in time the Hubble parameter $H$ increases and $\dot{H}>0$. We elaborate below the required properties of bouncing cosmological models,

\begin{itemize}
\item At the bouncing epoch, the scale factor $a$ contracts to non-zero finite size, the Hubble parameter $H$ to vanish and the deceleration parameter $q=-1+\frac{\dot{H}}{H^2}$ becomes singular.   

\item The NEC is violated, since the Hubble parameter changes sign from the bouncing point, therefore this phenomena is ruled out in the context of General Relativity (GR).

\item The slope of the scale factor increases after the bounce. During the matter contraction phase the Hubble parameter remains negative. It becomes positive during the matter expansion process. 
\end{itemize}
So, to frame the bouncing cosmological model with the above mentioned properties, we have considered the bouncing scale factor in the form, $a(t)=\left(\frac{\alpha}{\chi}+t^{2}\right)^{\frac{1}{2\chi}}$, where $\alpha$ and $\chi$ both are positive constants and subsequently, we obtain $H=\frac{t}{\alpha +\chi t^{2}}$ and $q=-1-\frac{\alpha}{t^{2}}+\chi$.\\

\begin{figure}[!htp]
\centering
\includegraphics[scale=0.50]{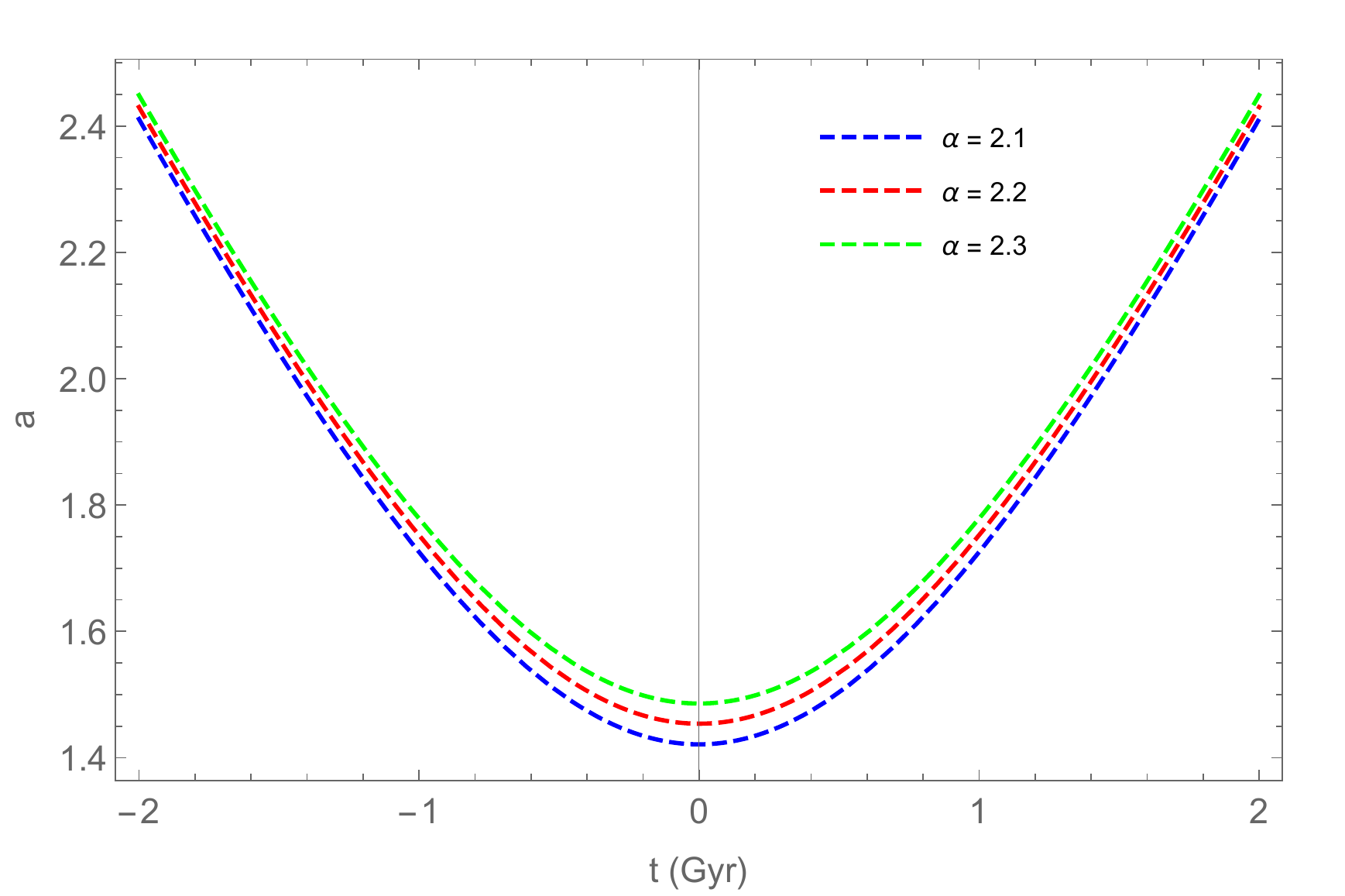}
\includegraphics[scale=0.50]{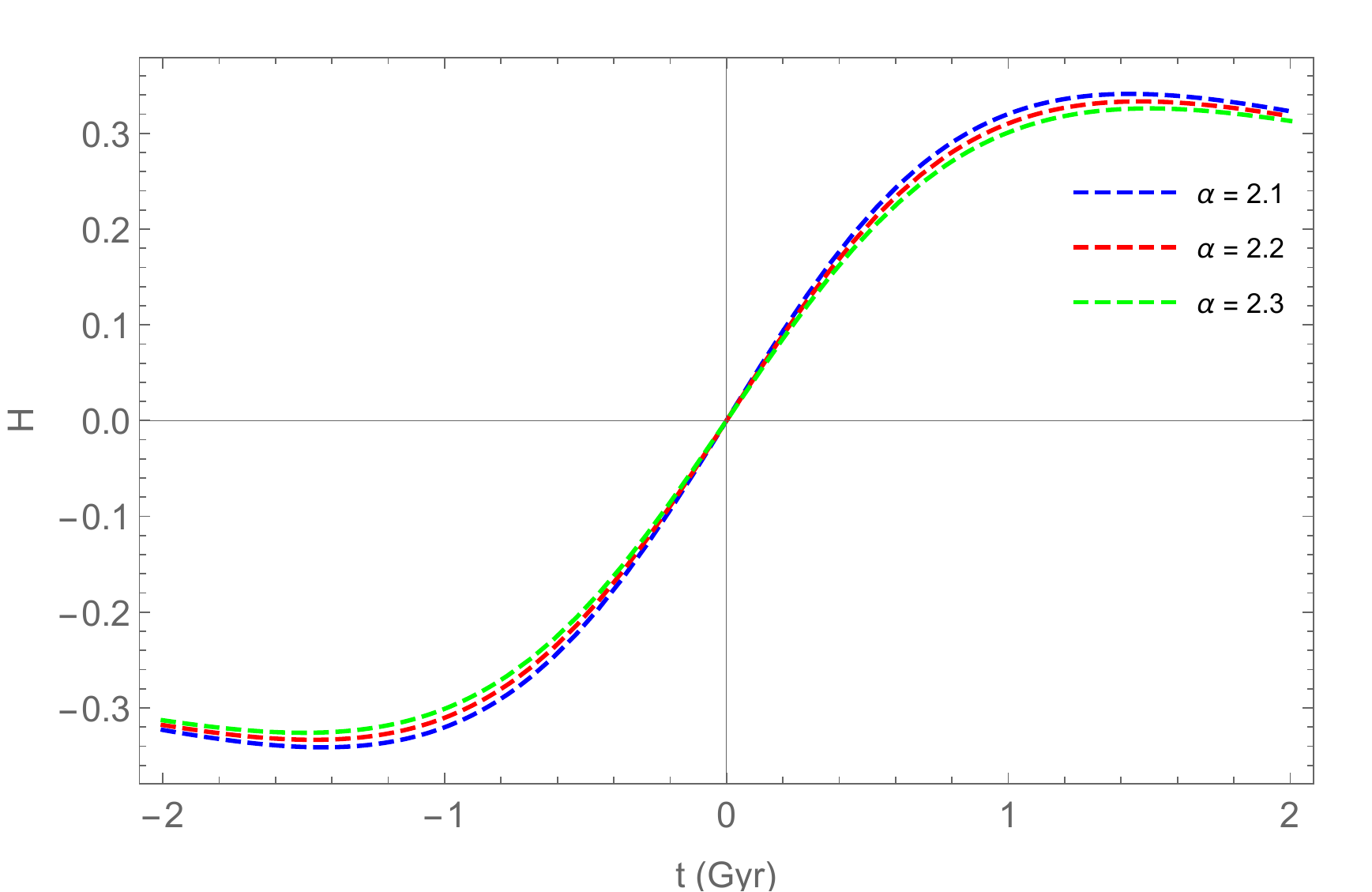}
\includegraphics[scale=0.50]{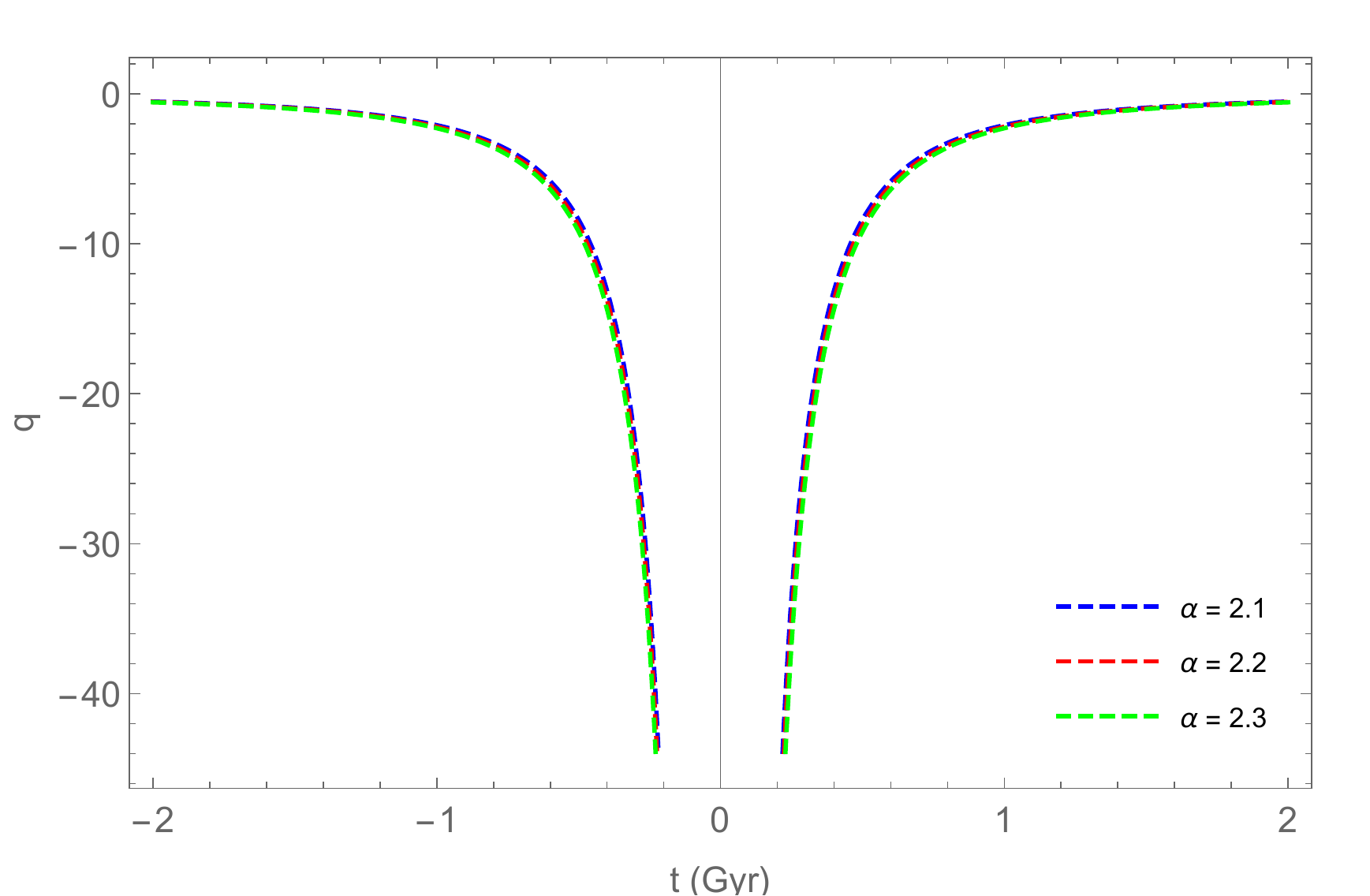}
\caption{ Evolution of scale factor ($a$) [left panel], Hubble parameter ($H$) [right panel] and deceleration parameter ($q$)  [below panel] in cosmic time $t$ for three representative values of the model parameter $\alpha$. The other parameter, $\chi=1.023$. The $x$-axis represents the cosmic time $t$.}
\label{Fig1}
\end{figure}

FIG. \ref{Fig1} represents the graphical behaviour of the scale factor, the Hubble parameter and the deceleration parameter. It may be observed that all of these quantities behave as per the properties prescribed in bouncing cosmology. The scale factor reduces from a higher value in early time to a small but finite value at the bounce and then increases after the bounce. The Hubble parameter starts from higher negative value, crosses the null point $H=0$ at bounce and then increases further. The deceleration parameter has been obtained as a function of the cosmic time. The graphical behaviour FIG. \ref{Fig1} (below panel) shows that the deceleration parameter evolves with time from a negative value $-1$ to an asymptotic value of $-1$ at late time. Also, it evolves both in pre and post bounce epoch. At the bouncing epoch, the deceleration parameter is experiencing the occurrence of singularity. Now, with this conducive behaviour of the basic parameters, we can proceed with the considered scale factor to frame the cosmological model and its analysis. We will adhere to analyse the dynamical parameters, the energy conditions, the cosmographic parameters and the other cosmological analysis. 

\subsection{Dynamical Parameters}
The constant deceleration parameter can be resulted in constant EoS and the time dependency of the deceleration parameter governs the time evolution of EoS parameter. Since in our model the deceleration parameter varies with time, here we shall discuss the behaviour of EoS parameter. For the bouncing scale factor  considered in the present work, the pressure, energy density and the EoS parameter can be obtained as, 

\begin{eqnarray}
p&=&-\frac{1}{(\eta^2-\beta^2)}\left[\frac{2\eta(\alpha-\chi t^2)+3(\eta-\beta)t^2}{(\alpha+\chi t^2)^2}\right]+\frac{\Lambda_0}{(\eta+\beta)}, \label{eq.13} \\
\rho&=& \frac{1}{(\eta^2-\beta^2)}\left[\frac{-2\beta(\alpha-\chi t^2)+3(\eta-\beta)t^2}{(\alpha+\chi t^2)^2}\right]-\frac{\Lambda_0}{(\eta+\beta)}, \label{eq.14} \\
\omega&=& -1+\left[\frac{2(\eta+\beta)(\alpha-\chi t^2)}{2\beta(\alpha-\chi t^2)-3(\eta-\beta)t^2+(\eta-\beta)(\alpha+\chi t^2)^2 \Lambda_0}\right]. \label{eq.15}
\end{eqnarray}

\begin{figure}[!htp]
\centering
\includegraphics[scale=0.50]{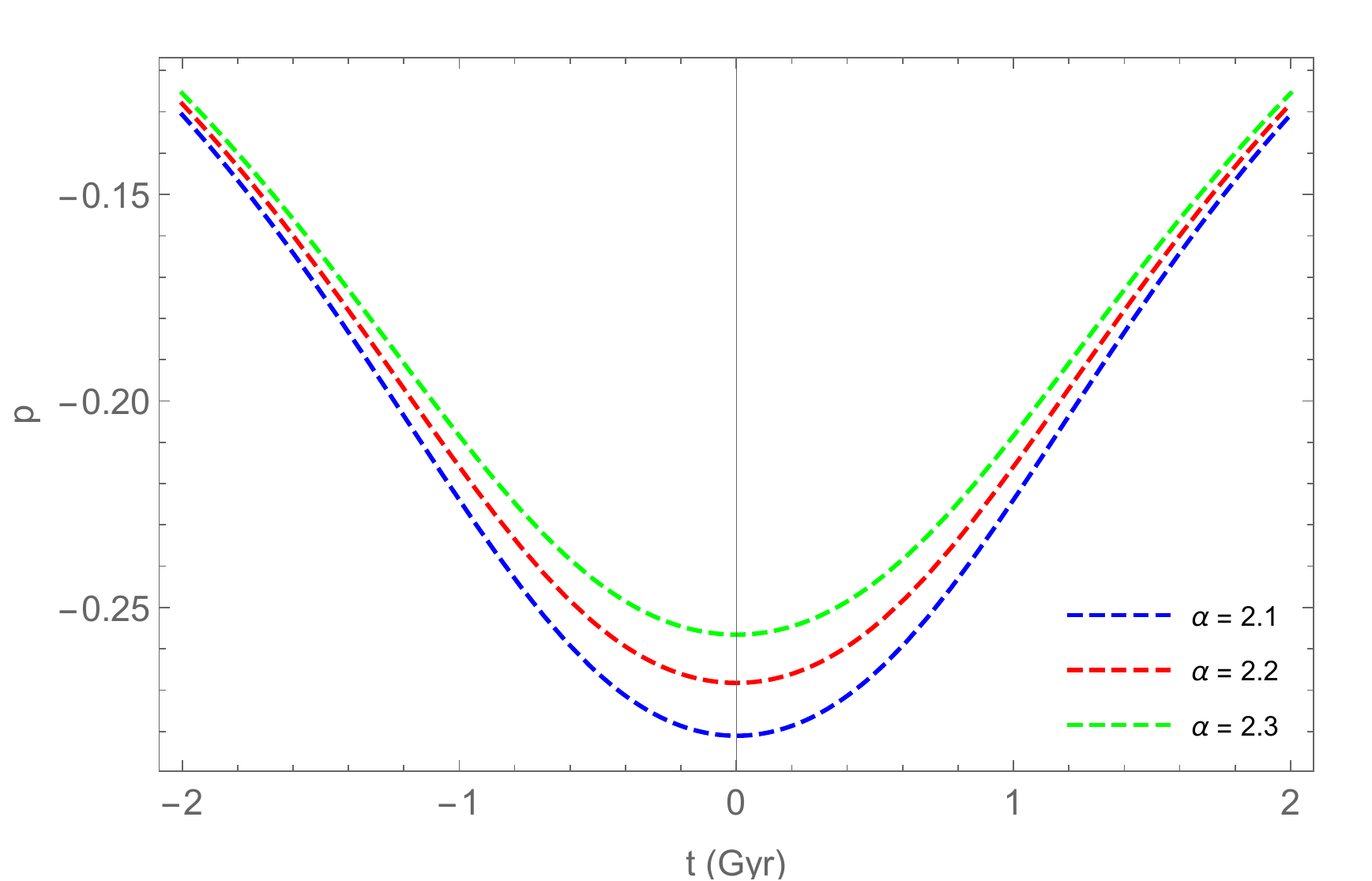}
\includegraphics[scale=0.50]{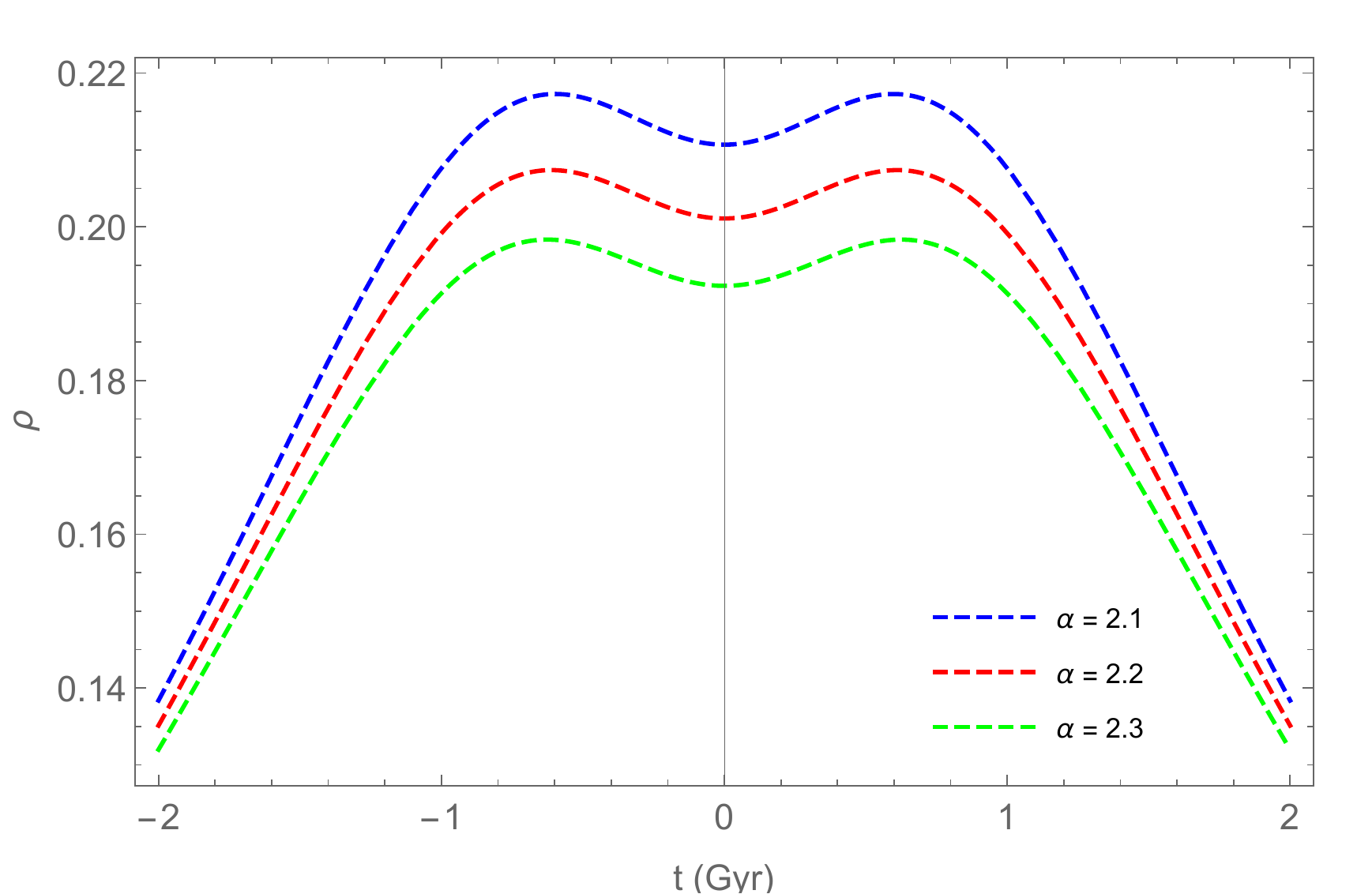}
\includegraphics[scale=0.50]{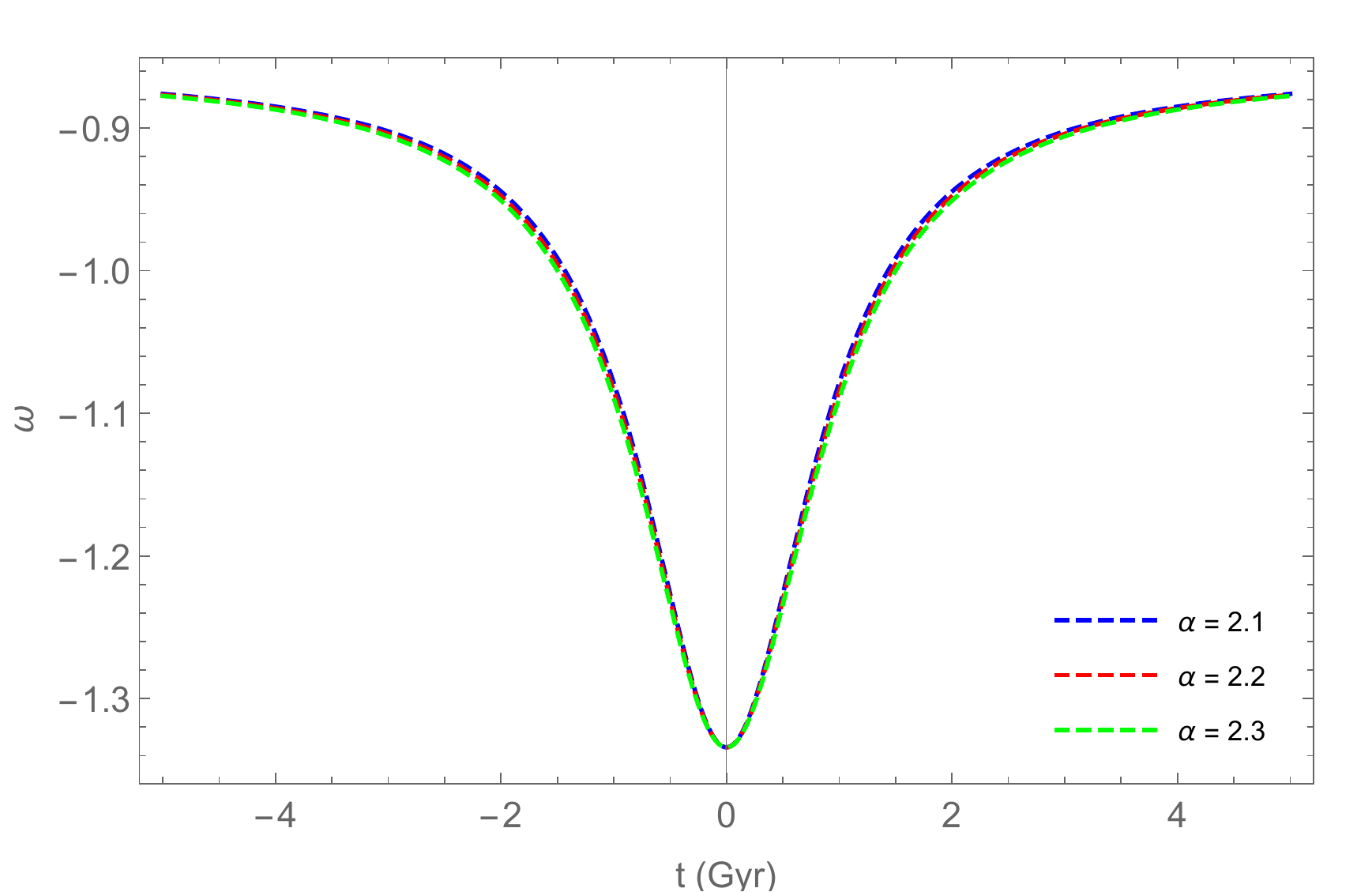}
\caption{Evolution of pressure ($p$) [left panel], energy density ($\rho$) [right panel] and EoS parameter ($\omega$) [below panel] in cosmic time $t$ for three representative values of the model parameter $\alpha$. The other parameters, $\chi=1.023$, $\beta=-5.8$, $\Lambda_0=0.001$. The $x$-axis represents the cosmic time $t$. }
\label{Fig2}
\end{figure}

The evolutionary behaviour of the matter pressure, energy density and EoS parameter have been represented graphically in FIG. \ref{Fig2}. The matter pressure shows the bounce at $t=0$ with a well-shaped for the three representative values of the model parameters $\alpha$ and remain in the negative profile throughout. The depth of the well decreases with the increase in the value of the parameter $\alpha$.  We have constrained the parameters to ensure that the energy density  remains positive at the bounce and it remains positive both in the negative and positive time zone. Also, it shows a bump near the bouncing epoch. In the contracting phase, the energy density increases sharply with the time nearing the bounce, attains a peak and then dips down at the bouncing epoch. After the bounce, it again increases for a short while and then decreases with an advancement of cosmic time. The slope of the energy density curve decreases with an increase in the value of the parameter $\alpha$. Also, at the bounce, the energy density dips to lower values for higher $\alpha$. In many similar earlier works, we have observed similar behaviour for the energy density near the bouncing epoch \cite{Tripathy2019, Tripathy21, Agrawal21}.  Possibly this behaviour of the energy density may be an essential feature to trigger a bounce followed by an expanding phase. The EoS parameter is obtained to be well-shaped near the bounce. We do not find any significant effect of the parameter $\alpha$ on the depth of the well. In fact, for all the values of $\alpha$ chosen in the present work, all the curves of the EoS parameter almost merge together near the bounce. However, in the tail region both in the negative and positive time domain, the curve of the EoS parameter marginally split for different values of $\alpha$. The EoS parameter, in both sides of the bouncing epoch evolves from the phantom region and when it moves away from the bouncing epoch the model pass through $\Lambda$CDM line and subsequently remain in the quintessence phase. Similar behaviour noticed in the evolution of EoS parameter in both the negative and positive time zone. 

\subsection{Energy Conditions}
By applying coordinate invariant restrictions on the total energy momentum tensor $T_{ij}$ the bouncing Universe can be ruled out and these restrictions are known as the energy conditions\cite{Novello08,Ashtekar11,Giovannini17}. In modified gravity, the bouncing solution should violate the NEC which in turn requires $T_{ij}k^{i}k^{j}\geq 0$, where $k^{i}$ be the null vector i.e $g_{ij}k^{i}k^{j}=0$. To address the weak energy condition (WEC), the energy momentum tensor should satisfy the inequality $T_{ij}u^{i}u^{j}\geq 0$ at each point of the space-time, where $u^{i}$ be the four velocity vector of the fluid, and the energy density of the given matter source should also be non-negative. According to the dominant energy condition(DEC), the matter flows along time like or null world lines and the strong energy condition (SEC) indicates that gravity to be attractive \cite{Raychaudhuri55,Ehlers06,Capozziello18}. Here we have presented the expressions of NEC, WEC, SEC, DEC respectively in the context of $f(R,T)$ gravity using eqns. \eqref{eq.13} and \eqref{eq.14} as,

\begin{eqnarray}
\rho+p &=& -\frac{2}{(\eta -\beta)} \left[\frac{\alpha-\chi t^2}{(\alpha+\chi t^2)^2}\right]\geq 0, \nonumber \\
\rho+p &=& -\frac{2}{(\eta -\beta)} \left[\frac{\alpha-\chi t^2}{(\alpha+\chi t^2)^2}\right]; \rho\geq 0, \nonumber \\
\rho+3p &=& \frac{1}{(\eta^2-\beta^2)}\left[\frac{(-2\beta-6\eta)(\alpha-\chi ^2t)-6(\eta-\beta)t^2}{(\alpha+ \chi t^2)^2}\right]+\frac{2\Lambda_0}{(\eta+\beta)}\geq 0,\nonumber \\
\rho-p&=& \frac{1}{(\eta^2-\beta^2)}\left[\frac{(-2\beta+2\eta)(\alpha-\chi ^2t)+6(\eta-\beta)t^2}{(\alpha+ \chi t^2)^2}\right]-\frac{2\Lambda_0}{(\eta+\beta)}\geq 0.
\end{eqnarray}
\begin{figure}[!htp]
\centering
\includegraphics[scale=0.50]{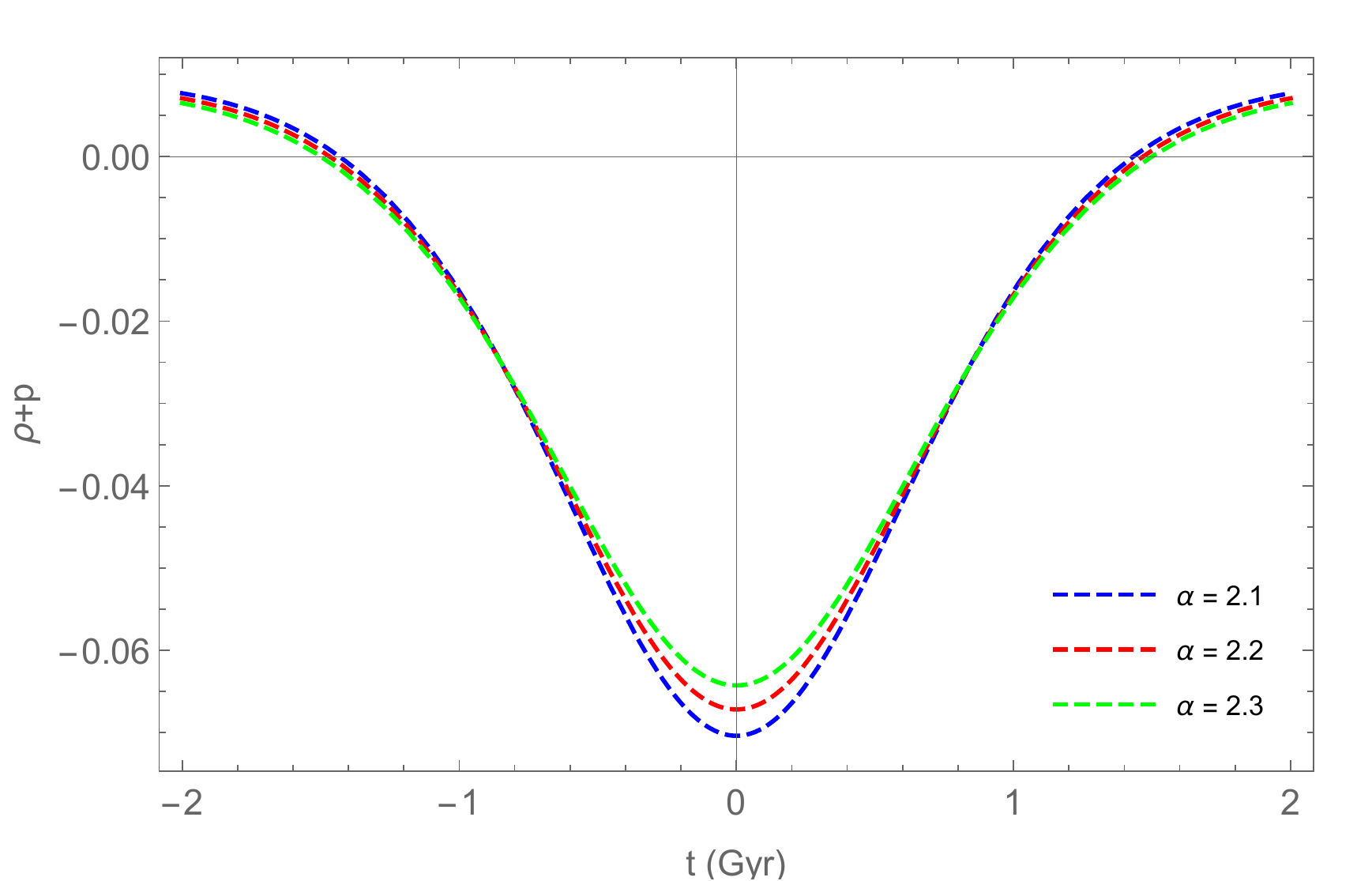}
\includegraphics[scale=0.50]{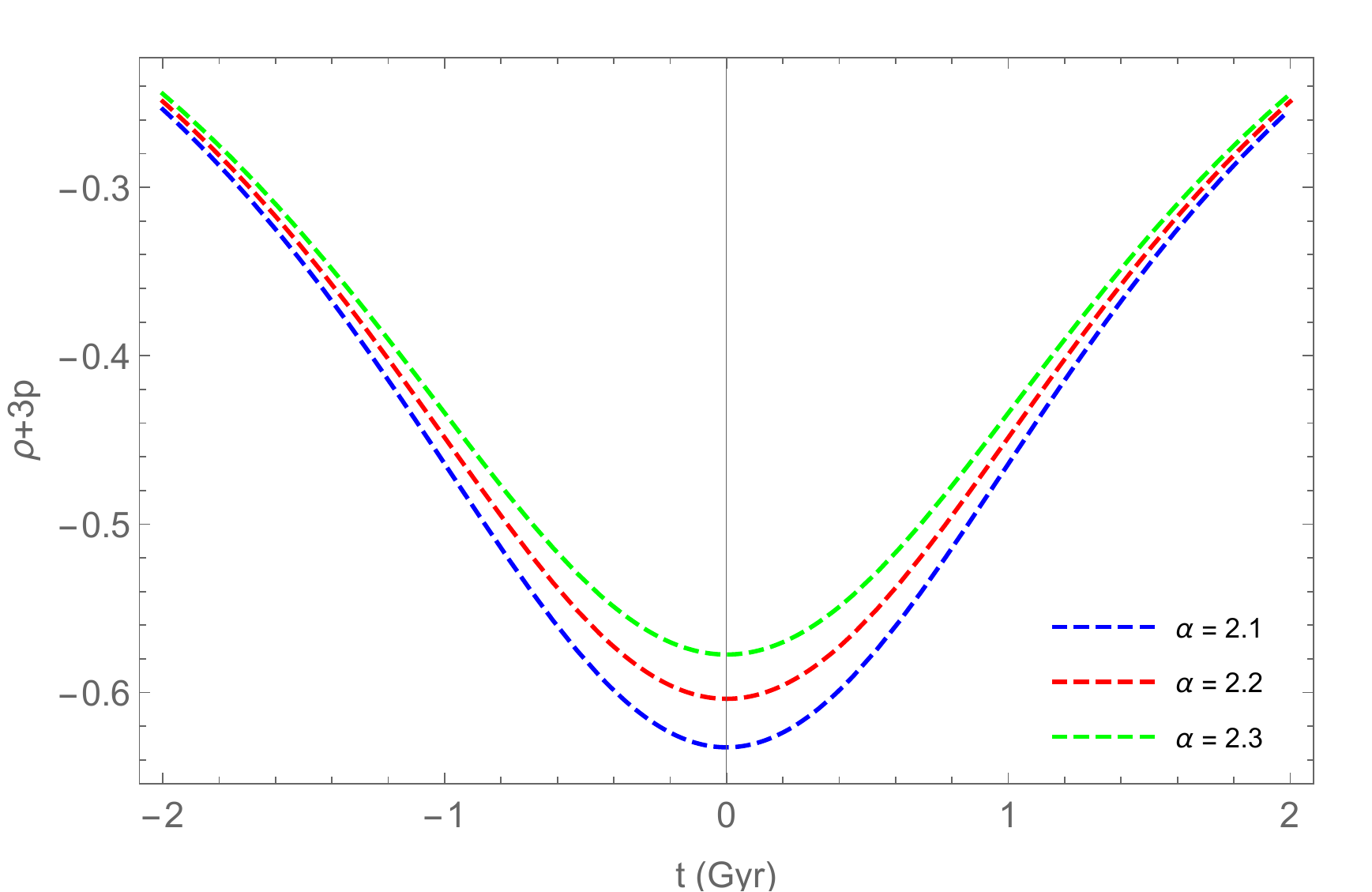}
\includegraphics[scale=0.50]{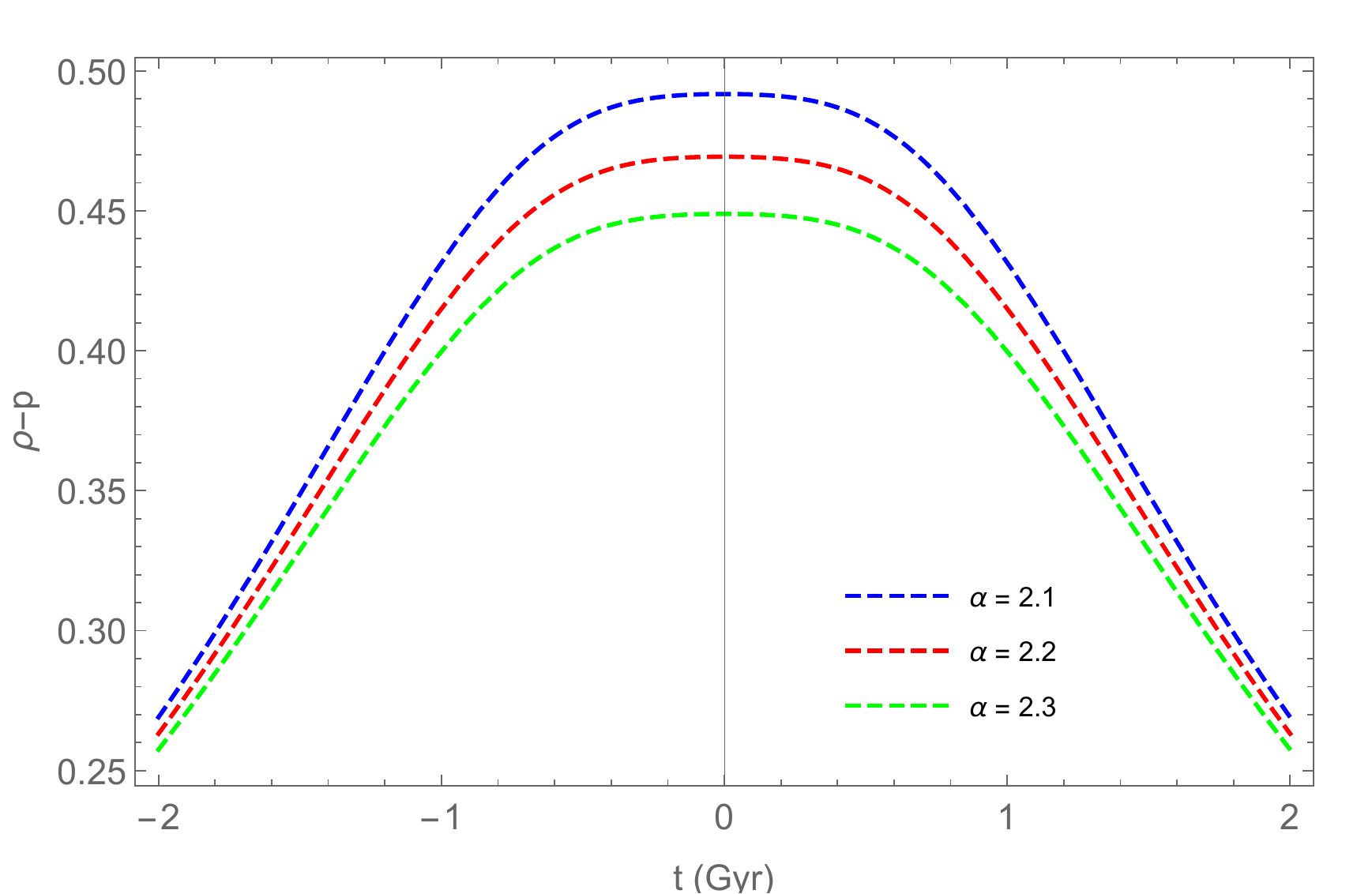}
\caption{Evolution of NEC/WEC ($\rho+p$) [left panel], SEC ($\rho+3p$) [right panel] and DEC ($\rho-p$)  [below panel] in cosmic time $t$ for three representative values of the model parameter $\alpha$. The other parameters, $\chi=1.023$, $\beta=-5.8$, $\Lambda_0=0.001$. The $x$-axis represents the cosmic time $t$.}
\label{Fig3}
\end{figure}
We have presented the graphical behaviour of energy conditions FIG. \ref{Fig3}. The bouncing behaviour demands the violation of NEC at the bouncing epoch, which can be easily observed in the left panel of FIG. \ref{Fig3}. The violation of NEC leads to the violation of SEC and it remains less importance in the study. But the DEC is expected to remain positive in the context of a perfect fluid and the same has been obtained. To reconstruct the extended gravity model, here we have used the bouncing scale factor in a simplified extended gravity, and in turn, through this reconstructed model, we can experience the bouncing scenario without invoking any dissipative fluid in the matter field. We considered only the contribution of a perfect fluid in a geometrically modified gravity theory to show that, bouncing scenario may be possible in the extended gravity theory. Also, we show that, the bouncing conditions are satisfied in the model along with the required violation/satisfaction of different energy conditions. 

\subsection{Cosmography}

Cosmography is the mathematical framework of the cosmological parameters to describe the Universe. Visser \cite{Visser04} has extended the idea of cosmography which was originally mentioned by Weinberg \cite{Weinberg72}. It depends on the Copernican principle leads to the FLRW metric. The cosmological parameters such as Hubble parameter, $H$; deceleration parameter, $q$; jerk parameter, $j$; snap parameter, $s$ and lerk parameter, $l$ constitute the cosmographic set. All these parameters are respectively associated with the increasing order of derivative of the scale factor and these can be obtained by the expansion of the scale factor with the cosmic time \cite{Alam03,Sahni03,Capozziello06, Capozziello08,Aviles11,Salehi17, Tripathy2019, Tripathy2020}. The Hubble parameter and the deceleration parameter are defined and analysed at the beginning of Section III above and here we shall derive the other cosmographic parameters as, 

\begin{eqnarray}
j&=&\frac{(2 \chi -1) \left[t^2 (\chi -1)-3 \alpha \right]}{t^2}, \nonumber \\
s&=&-\frac{(2 \chi -1) \left[3 \alpha ^2+t^4 (\chi -1) (3 \chi -1)+6 \alpha  t^2 (1-3 \chi )\right]}{t^4}, \nonumber\\
l&=& \frac{\left(8 \chi ^2-6 \chi +1\right) \left[15 \alpha ^2+t^4 (\chi -1) (3 \chi -1)+10 \alpha  t^2 (1-3 \chi )\right]}{t^4}.
\end{eqnarray}

\begin{figure}[!htp]
\centering
\includegraphics[scale=0.50]{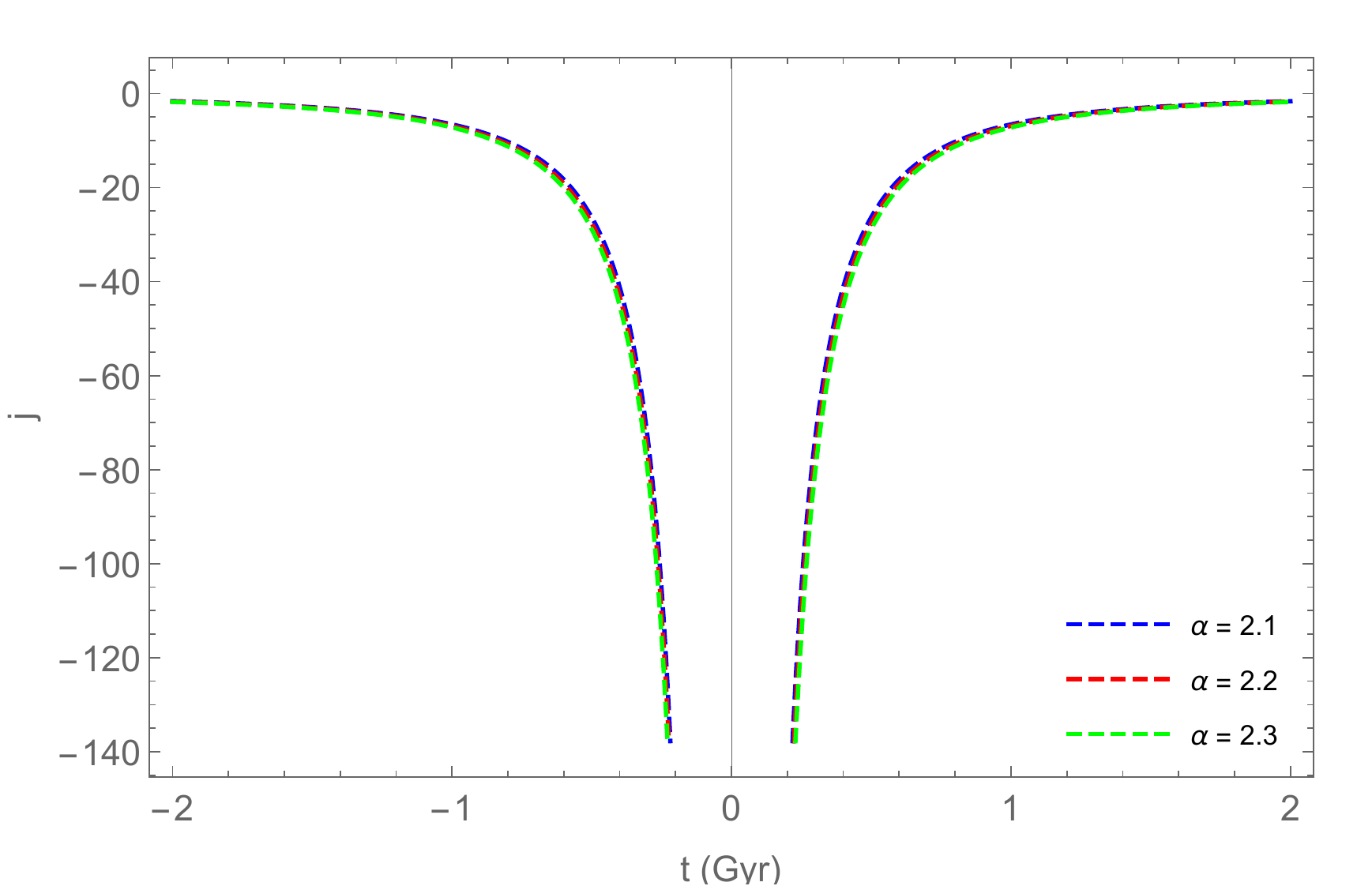}
\includegraphics[scale=0.50]{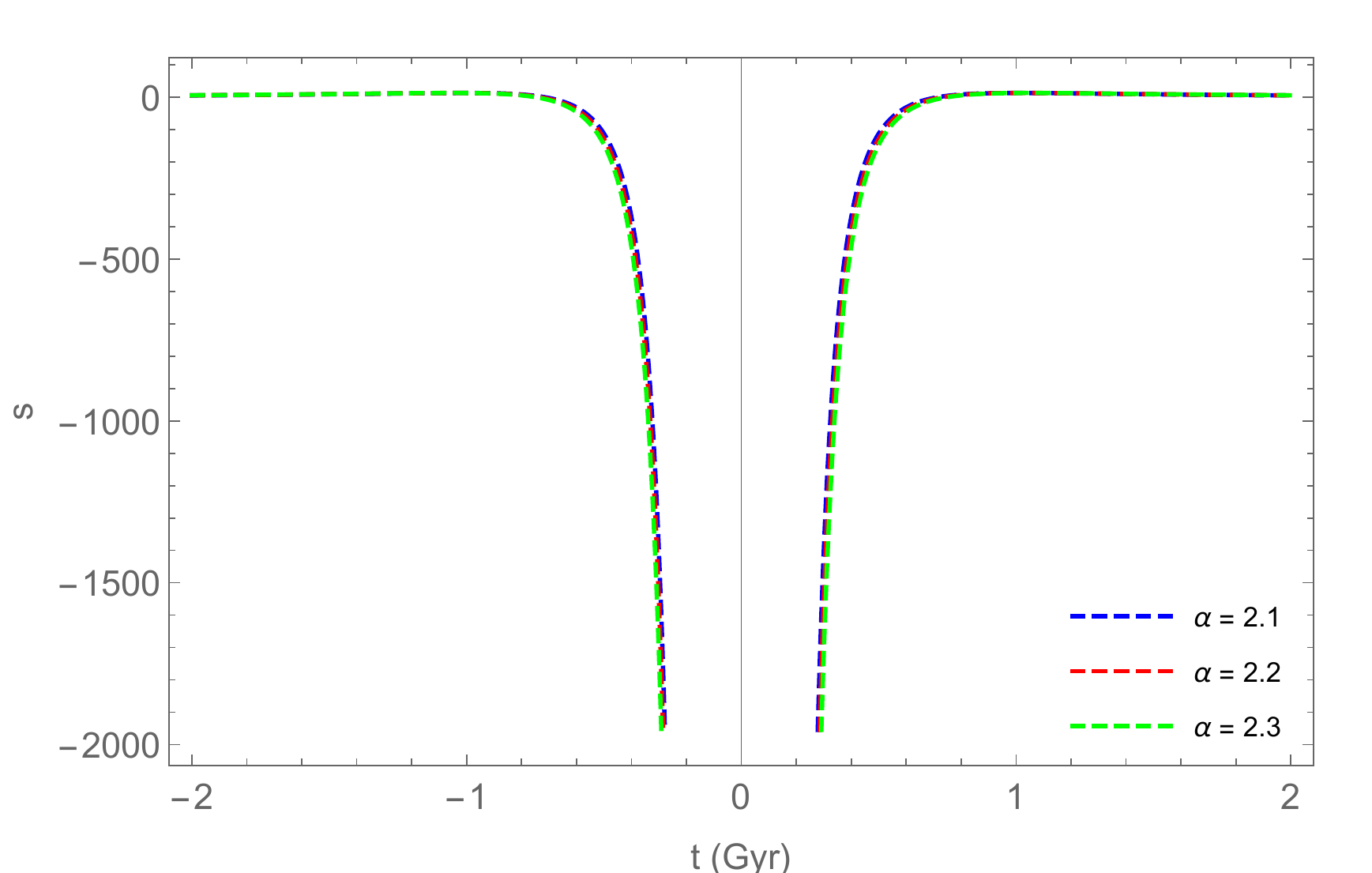}
\includegraphics[scale=0.50]{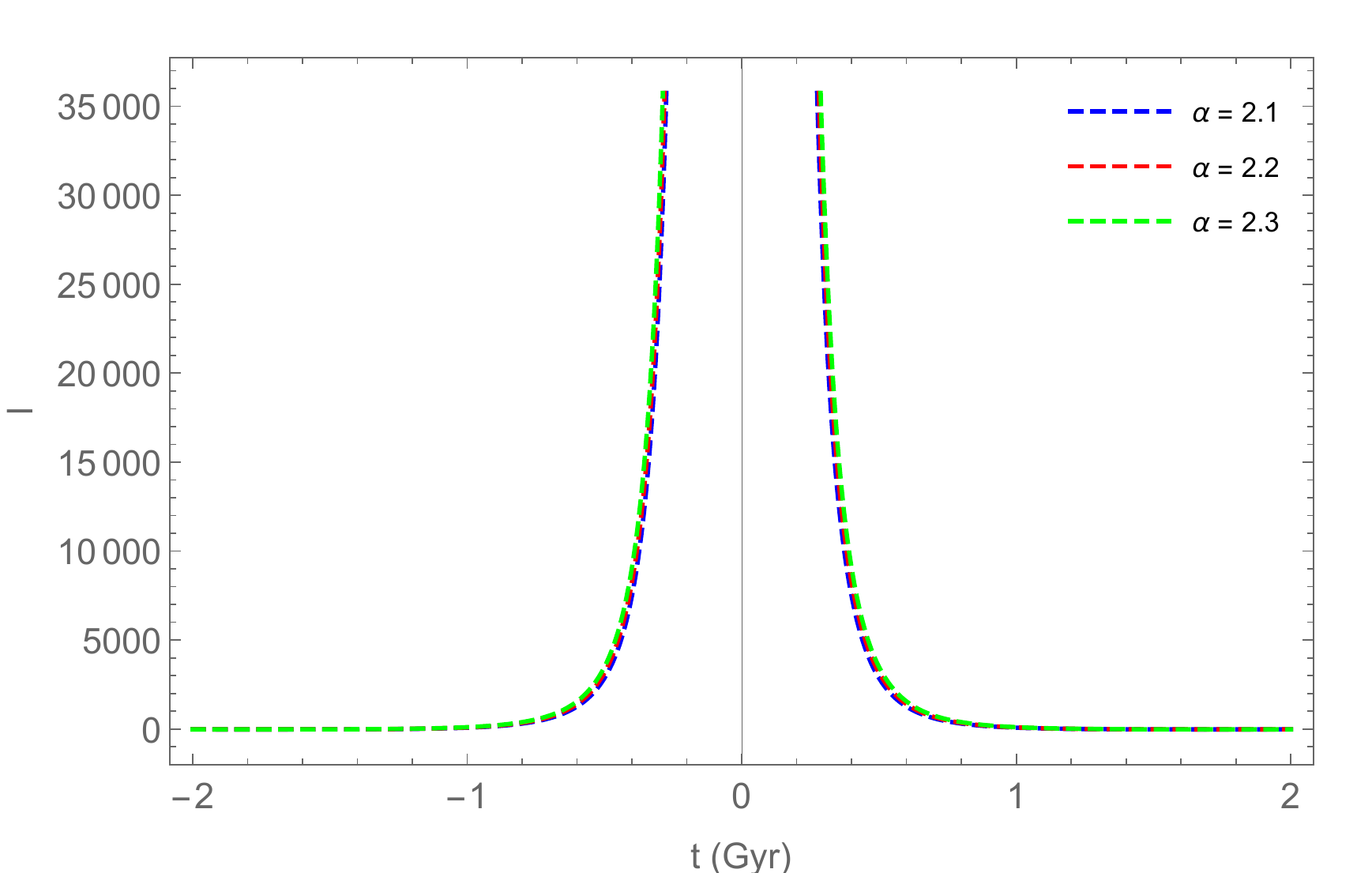}
\caption{Evolution of jerk parameter ($j$) [left panel], snap parameter ($s$) [right panel] and lerk parameter ($l$) [below panel] in cosmic time $t$ for three representative values of the model parameter $\alpha$. The other parameter, $\chi=1.023$. The $x$-axis represents the cosmic time $t$.}
\label{Fig4}
\end{figure}

The graphical behaviour of these cosmographic parameters have been presented in FIG. \ref{Fig4}. We have already seen that the EoS parameter evolves from the phantom region and as we move away from the bouncing epoch, the EoS parameter stays mostly in the quintessence region. Now the jerk parameter, which involves the third derivative of the scale factor shows a singular bounce at the bouncing epoch, however, it evolves from the $\Lambda$CDM behaviour ($j=1$) in both the negative and positive time zone. Similar behaviour has also been obtained for the snap parameter, which involves the fourth derivative of the scale factor, except the fact that  it evolves from its prescribed $\Lambda$CDM behaviour ($s=0$). At the same time the lerk parameter remains positive in both the time zone and evolve from the matter dominated Universe.

\subsection{Stability of the Model}
\begin{figure}[!htp]
\centering
\includegraphics[scale=0.50]{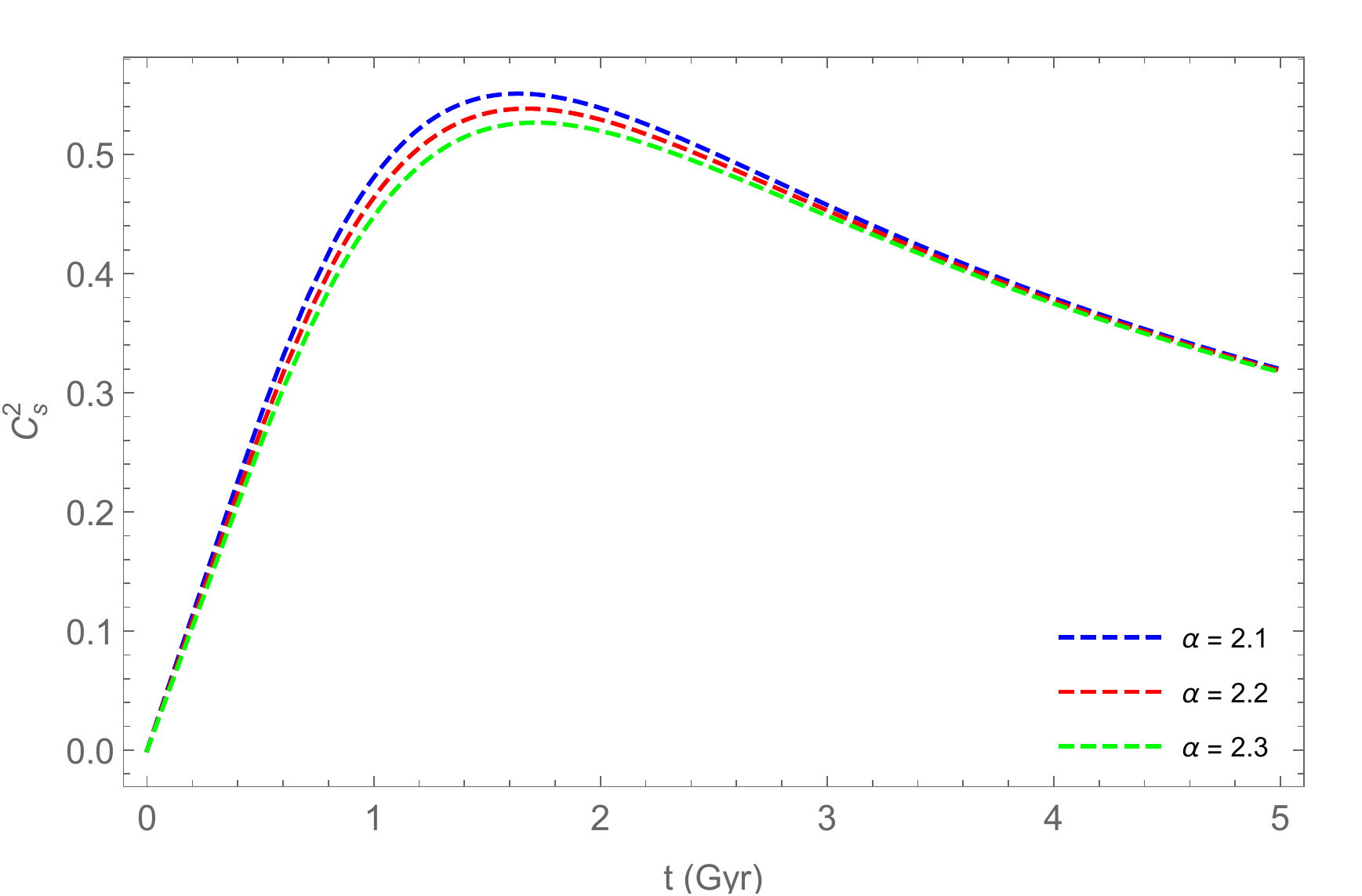}
\caption{Evolution of stability analysis ($C_s^2$) in cosmic time $t$ for three representative values of the model parameter $\alpha$. The other parameters, $\chi=1.023$, $\beta=-5.8$, $\Lambda_0=0.001$. The $x$-axis represents the cosmic time $t$.}
\label{Fig5}
\end{figure}

It is important to test the stability of the cosmological model since several assumptions are being taken to solve the systems and the difficulty in assessing the degree of the generality in these assumptions. The stability of the cosmological model can be performed by calculating the adiabatic speed of sound through the cosmic fluid, $C_s^2=dp/d\rho$ \cite{Sudarsky95,Charters01,Farajollahi11,Mishra21}. The model can be termed as stable or unstable, when $C_s^2>0$ or $C_s^2<0$ respectively. Using eqns. \eqref{eq.13}-\eqref{eq.14}, we can obtain the stability relation of the model  in terms of cosmic time as,

\begin{equation}
C_{s}^{2}= \frac{6 \alpha  t (\beta +\eta  (2 \chi -1))-2 t^3 \chi  (3 \beta +\eta  (2 \chi -3))}{\left(\alpha +t^2 \chi \right) \left(\alpha  (\alpha  \lambda  (\beta -\eta )-2 \beta )+t^2 (2 \chi  (\alpha  \lambda  (\beta -\eta )+\beta )-3 \beta +3 \eta )+\lambda  t^4 \chi ^2 (\beta -\eta )\right)}.
\end{equation}
The graphical analysis for the  stability of  the model is shown in FIG. \ref{Fig5}. It may be inferred from the figure that, the model remains stable throughout the evolution for the representative values of the parameters. The stability graph shows a steep increase in the value of $C_s^2$ at the beginning and after $t=1.6$ it decreases slowly. Since the model shows the bouncing behaviour at the epoch $t=0$, and the stability range includes the bouncing epoch, we conclude that the presented model is stable. Also, for the present model, we achieve mechanical stability followed by the avoidance of supraluminar behaviour of the cosmic fluid. Interestingly, during the transition from the contracting phase to the expanding phase through the bounce, we have the EoS parameter $\omega <-1$ for the different choice of the model parameter $\alpha$. This is in contradiction to the EoS parameter obtained in models with slow contraction where $\omega >1$. It is worth to mention here that, in slow contraction process, a scalar field with a steep negative potential is considered as a bouncer field whose energy density redshifts faster than $a^{-6}$ leading to the homogenization of the Universe. The anisotropic stress remains subdominant and this consequently provides a solution to the flatness and horizon problem.

In addition to the above discussion, we wish to study the stability of the presented model against homogeneous and isotropic linear perturbation. We consider a pressureless dust FRW background whose general solution may be $H(t)=H_b(t)$. Following Ref.\cite{Dombriz2012}, we consider perturbations of the Hubble parameter and the energy density around the arbitrary solutions $H_b(t)$ as
\begin{eqnarray}
H(t) &=& H_b\left(1+\delta (t)\right),\\
\rho (t) &=& \rho_b\left(1+\delta_m (t)\right),
\end{eqnarray}
where $\delta_m(t)$ and $\delta(t)$ are the respective deviations from the background energy density and the Hubble parameter. In the present extended gravity theory, we have considered the functional $f(R,T)=R+2\beta T+2\Lambda_0$ which may be expanded in powers of $R_b$ and $T_b$ as
\begin{equation}
f(R,T)=f_b+(R-R_b)+2\beta(T-T_b)+\mathcal{O}^2,
\end{equation}
where the term $\mathcal{O}^2$ includes all the higher powers of $R$ and $T$. 

Using the perturbative approach in the equivalent FRW equation, we obtain
\begin{equation}
6H_b^2\delta(t)=\eta \rho_b\delta_m(t),\label{eq:22}
\end{equation}
which establishes an algebraic relationship between the geometry and matter perturbations. Here we have 
\begin{equation}
\rho_b=\frac{3H_b^2-\Lambda_0}{\eta+\beta}.
\end{equation}
From the conservation equation, we may obtain 
\begin{equation}
\dot{\delta}_m(t)+3H_b(t)\delta(t)=0.\label{eq:24}
\end{equation}

From eqs. \eqref{eq:22} and \eqref{eq:24}, the evolution equation for the perturbation in the Hubble parameter may be obtained as
\begin{equation}
\dot{\delta}(t)+\frac{\eta\rho_b}{2H_b}\delta(t)=0.\label{eq:25}
\end{equation}
For a vanishingly small value of $\Lambda_0$, the evolution equation \eqref{eq:25} reduces to
\begin{equation}
\dot{\delta}(t)+\frac{3\eta H_b}{2\left(\eta+\beta\right)}\delta(t)=0.\label{eq:26}
\end{equation}

For a bouncing scenario as prescribed in Section \ref{sect-iii}, eq.\eqref{eq:26} may be integrated to obtain the geometrical perturbation as
\begin{equation}
\delta(t)=C_k\left(\alpha+\chi t^2\right)^{-\frac{3\eta}{4\chi\left(\eta+\beta\right)}},\label{eq:27}
\end{equation}
where $C_k$ is an integration constant.

\begin{figure}[t]
\centering
\includegraphics[scale=0.50]{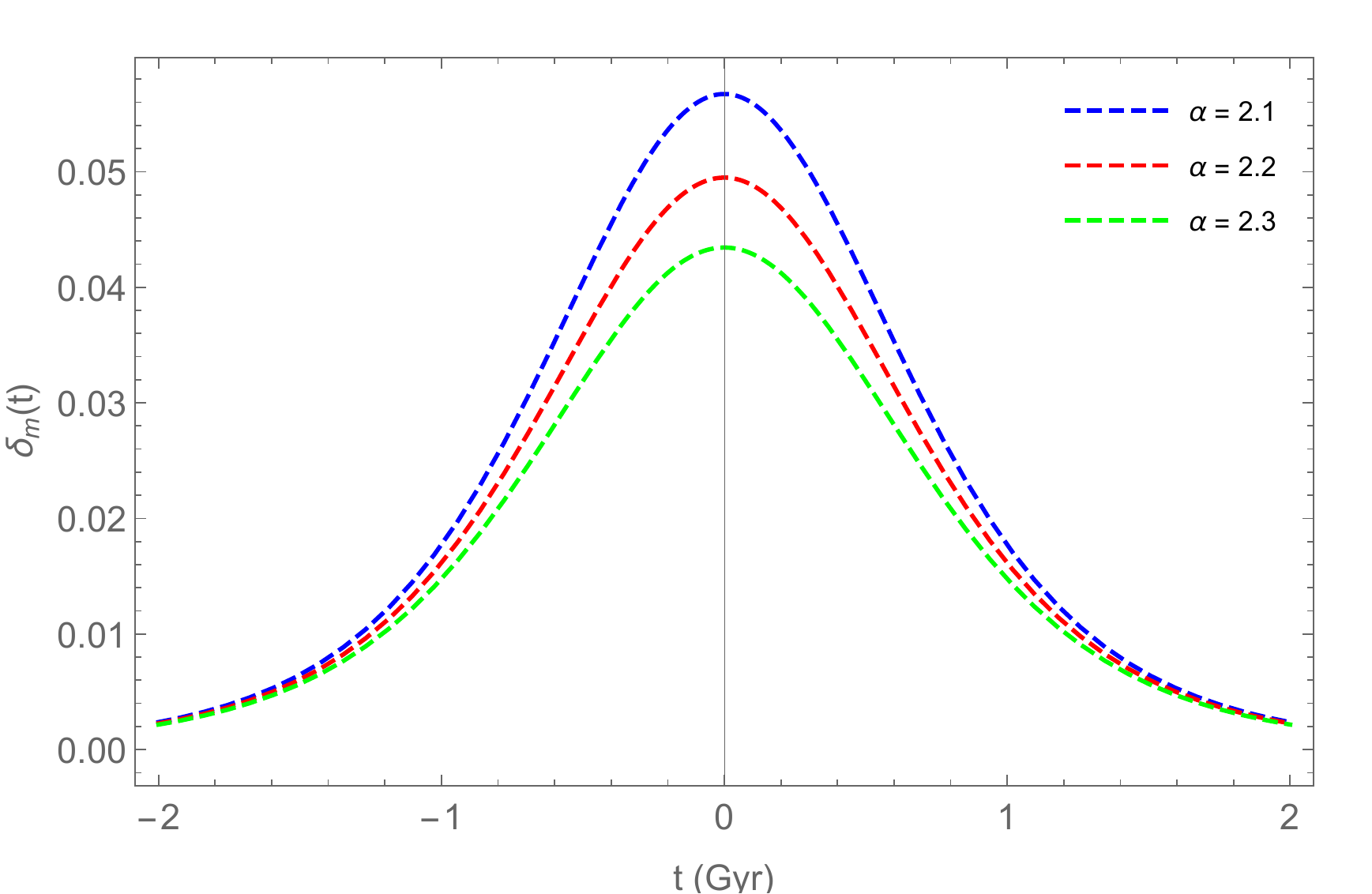}
\includegraphics[scale=0.50]{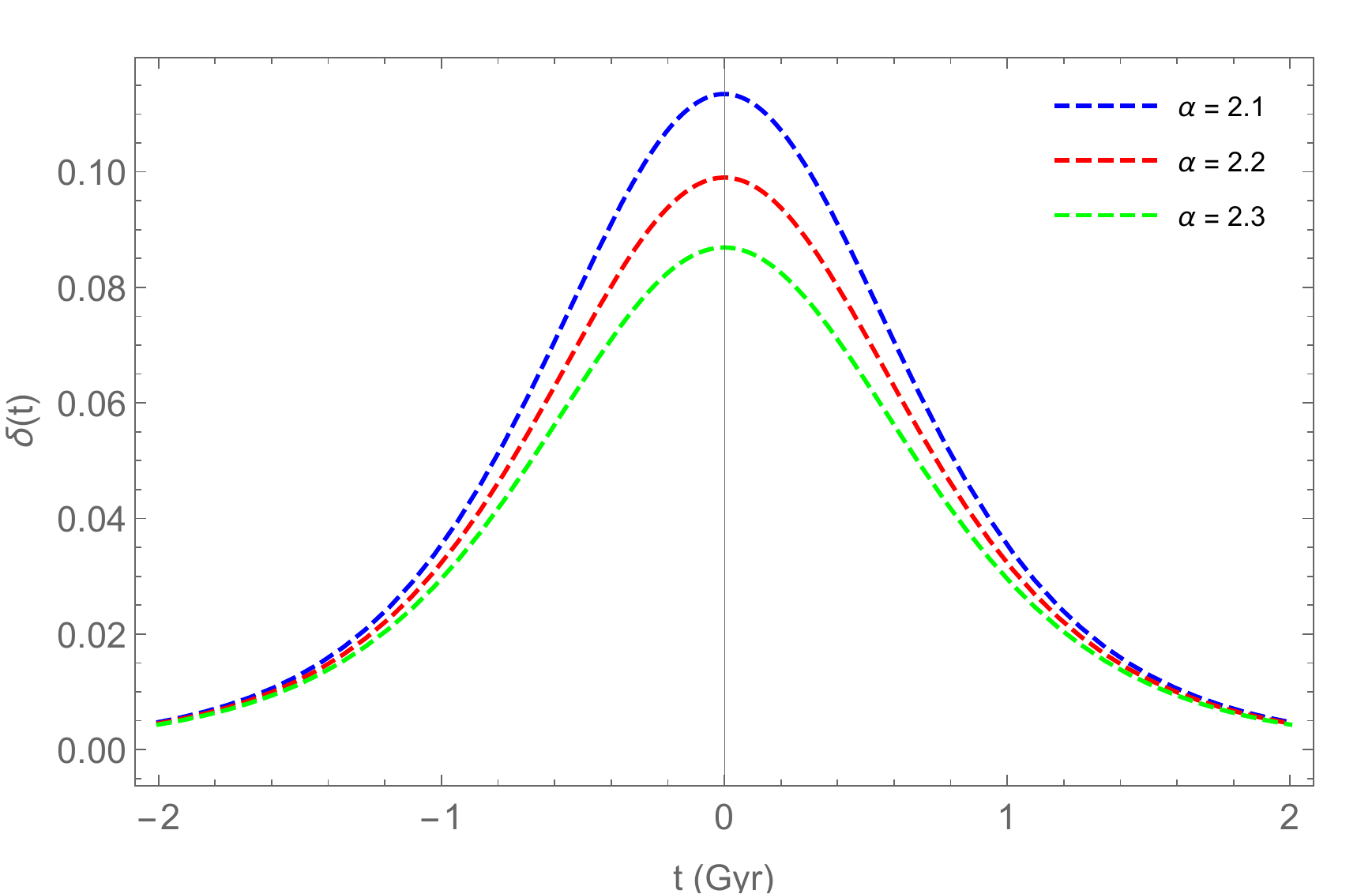}
\caption{Evolution of the perturbations in the energy density ($\delta_m$) [left panel] and the Hubble parameter ($ \delta$) [right panel] for three representative values of the model parameter $\alpha$. The other parameter space remains the same as mentioned earlier.}
\label{Fig6}
\end{figure}

The matter perturbation may be obtained in a straightforward manner from eq.\eqref{eq:22} as
\begin{equation}
\delta_m(t)=\frac{2C_k\left(\eta+\beta \right)}{\eta}\left(\alpha+\chi t^2\right)^{-\frac{3\eta}{4\chi\left(\eta+\beta\right)}}.\label{eq:28}
\end{equation}

In view of the above discussion, we may infer that, the geometric and matter perturbations $\delta(t)$ and $\delta_m(t)$ behave as $a^{-\frac{\eta}{2(\eta+\beta)}}$. In the present work, we chose the parameters $\eta$ and $\eta+\beta$ as positive quantities, therefore, the linear perturbations in the Hubble parameter and the energy density symmetrically decrease with the growth of time as we move away from the bouncing epoch. We show the behaviour of these linear homogeneous perturbations for three different values of the model parameters in FIG. 6. The gradual decreasing behaviour of the perturbations clearly articulates the stability of the model under a bouncing scenario.
\section{Reconstruction by redshift parameter }
In cosmology the cosmic time ($t$) related to the cosmological red shift ($z$) as, $t\approx\frac{14Gyr}{(1+z)^{\frac{3}{2}}}$ and also the red shift and scale factor can be related as, $\frac{a(t)}{a_{0}}=\frac{1}{1+z}$, where $a_{0}$ is the value of scale factor at the present time considered to be one. Here we wish to show the behaviour of the dynamical parameters with that of cosmological redshift. We shall further assume a dimensionless parameter $r(z)=\frac{H^2(z)}{H_0^2}$, where $H_0$ be the present value of the Hubble parameter with $H_0=71\pm 3  kms^{-1}Mpc^{-1}$. With the above, we can have the differential form of scale factor and red shift as, 
\begin{equation}
\frac{d}{dt}=\frac{da}{dt}\frac{dz}{da}\frac{d}{dz}=-H(z+1)\frac{d}{dz}.   
\end{equation}
Now, the dynamical parameters eqs. \eqref{eq.10}-\eqref{eq.12} can be rewritten as a redshift function as, 
\begin{eqnarray}
p&=& -\frac{1}{(\eta^2-\beta^2)}\left[-\eta(z+1)H_{0}^2 r'(z)+3(\eta-\beta)H_{0}^2 r(z)-(\eta-\beta)\Lambda_0\right], \label{eq.19} \\ 
\rho&=& \frac{1}{(\eta^2-\beta^2)}\left[\beta (z+1)H_{0}^2 r'(z)+3(\eta-\beta)H_{0}^2 r(z)-(\eta-\beta)\Lambda_0 \right], \label{eq.20}\\ 
\omega&=&-1+\left[\frac{-(\eta+\beta) (z+1)H_{0}^2 r'(z)}{-\beta (z+1)H_{0}^2 r'(z)-3(\eta-\beta)H_{0}^2 r(z)+(\eta-\beta)\Lambda_0}\right],\label{eq.21}
\end{eqnarray}
where prime denotes derivative with respect to red shift.
\begin{figure}[!htp]
\centering
\includegraphics[scale=0.50]{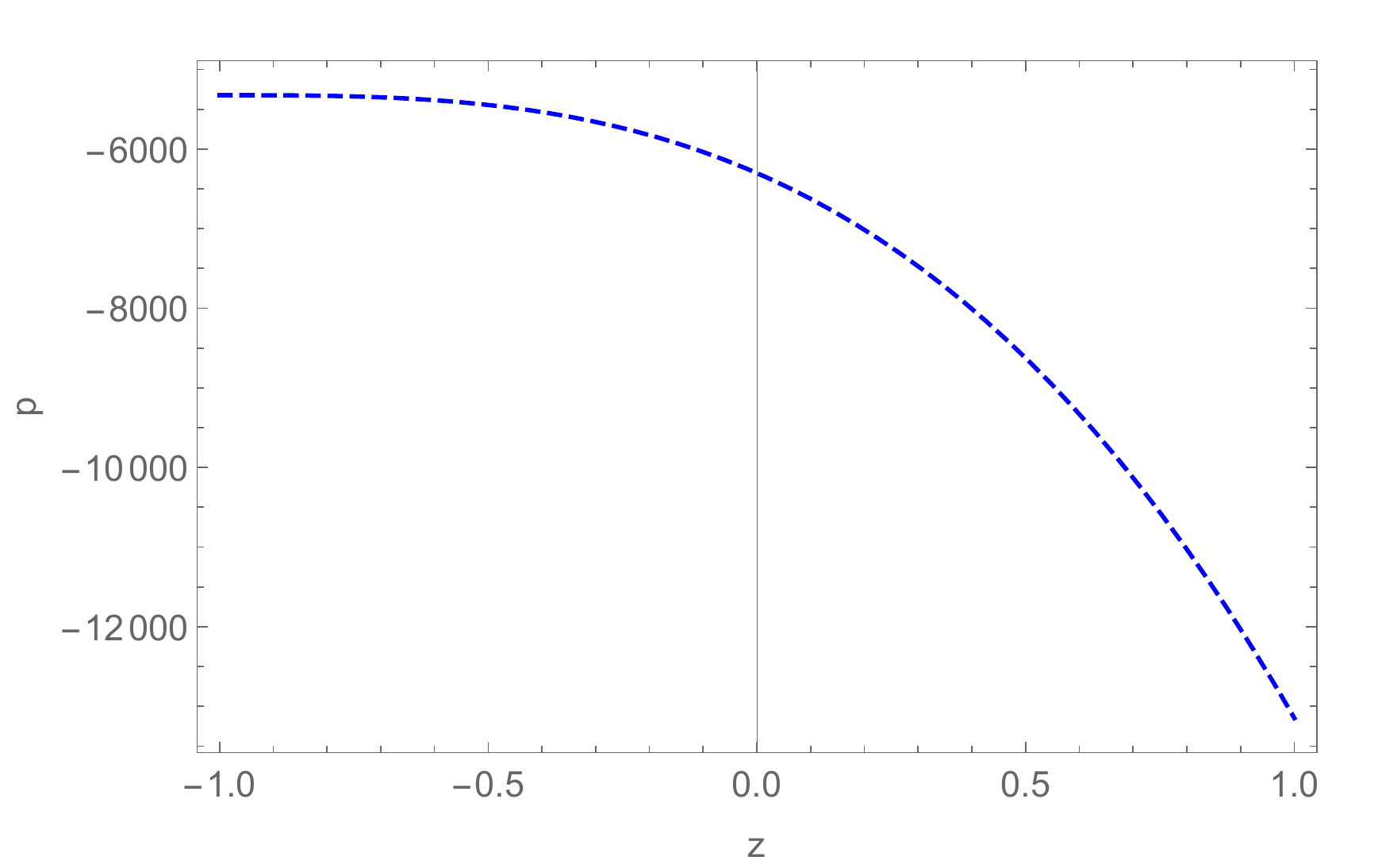}
\includegraphics[scale=0.50]{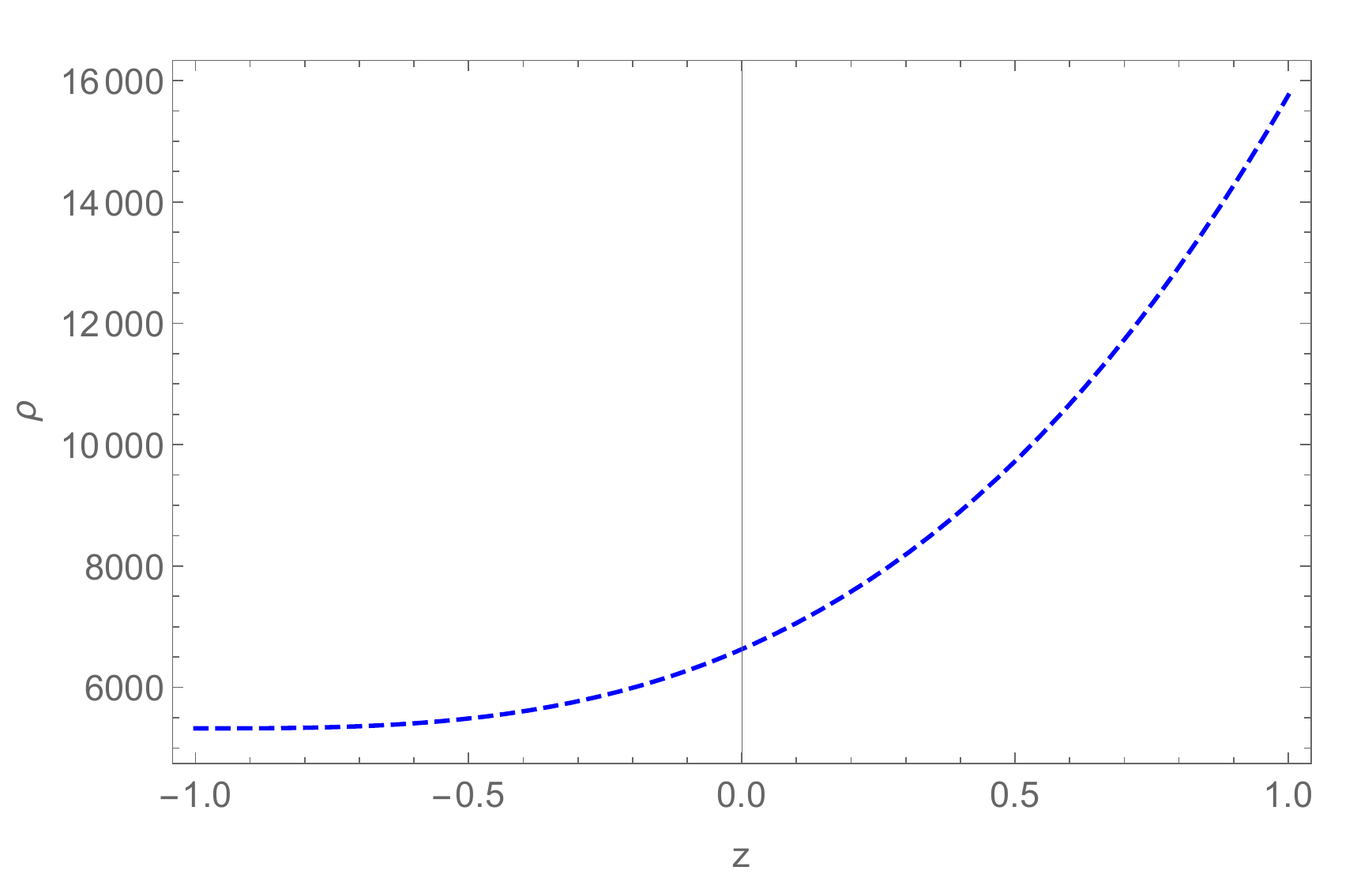}
\includegraphics[scale=0.50]{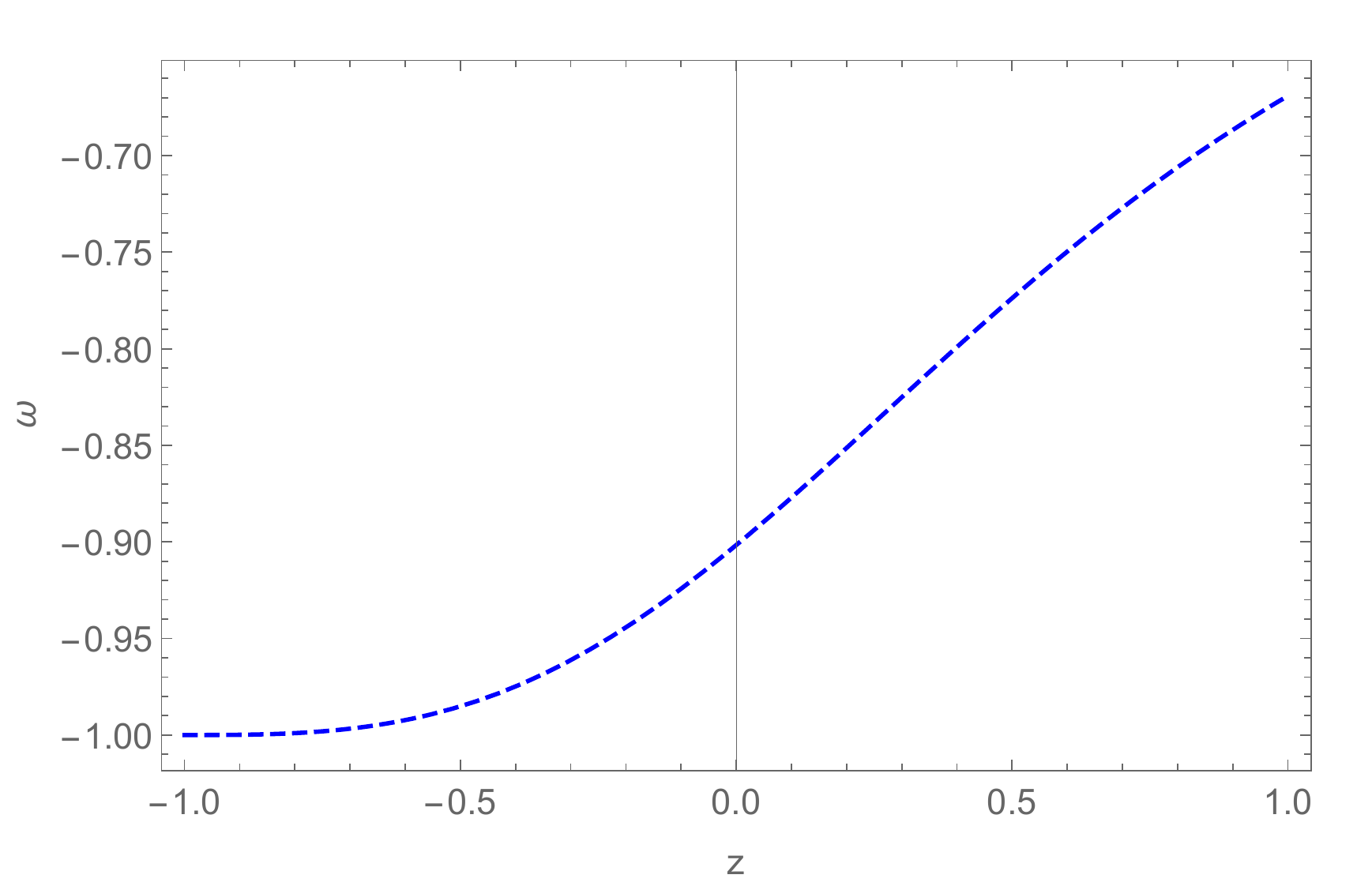}
\caption{Evolution of pressure ($p$) [left panel], energy density ($\rho$) [right panel] and EoS parameter ($\omega$) [below panel] in redshift $z$. The parameters, $\Lambda_0=0.001$, $\beta=-5.8$. The $x$-axis represents the redshift $z$.}
\label{Fig7}
\end{figure}

We have incorporated the dimensionless parameter, $r(z)$ to reconstruct the model as the dark energy model.  The energy density of the matter $\rho_m$, the EoS parameter $\omega_m$ and redshift $z$ can be related as,
\begin{equation}
\rho_m=\rho_{m0}\left(\frac{a_0}{1+z}\right)^{-3(\omega_m+1)}
\end{equation} 
We shall reconstruct the dark energy model with the expression of $r(z)$ as \cite{Alam04,Copeland06,Ilyas21},
\begin{equation}\label{eq.22}
r(z)=C_{0}+C_{1}(1+z)+C_{2}(1+z)^{2}+\Omega_{m0}(1+z)^{3} ,   
\end{equation}
where, $C_{0}+C_{1}+C_{2}+\Omega_{m0}=1$. From the previous discussion, we have seen that the model is showing the $\Lambda$CDM behaviour and the parametrization considered here is based on $\Lambda$CDM scenario, hence the value of the constants plays an important role to connect the model and the parametrization, we obtained the value, $C_{0}=0.70,C_{1}=0,C_{2}=0$ and $\Omega_{m0}=0.30$. Along with these calculated values, we take the value of $H_0=70$ and have plotted the dynamical parameters in FIG. \ref{Fig7}.  We can observe the similar behaviour as expected the pressure remains entirely in negative domain, the energy density in positive domain and the EoS parameter shows $\Lambda$CDM behaviour at late time of the evolution. 

\section{Conclusion}
We have discussed the bouncing cosmological model of the Universe in $f(R,T)$ gravity with the matter field in the form of perfect fluid. From the behaviour of the dynamical parameters, we have noticed the bounce at the epoch $t=0$. The Hubble parameter crosses $H=0$ at the bouncing epoch thereby confirms the occurrence of cosmological bounce. Another criteria for the bouncing scenario is the violation of NEC and we have obtained the violation of NEC in the range where the bounce occurs. The behaviour of energy density and the EoS parameter respectively support the bouncing behaviour and remain entirely in positive and negative region respectively. At the bounce, since $\omega<-1$, the model is experiencing phantom field like behaviour, however when it evolves out, passes through $\Lambda$CDM line and subsequently to the quintessence phase. The behaviour of the deceleration parameter at the bounce shows singularity as in the case of other cosmographic parameters such as the jerk parameter, snap parameter and lerk parameter. The interesting feature of the bouncing model is that both jerk and snap parameter attain the singularity at its negative profile where as the lerk parameter in its positive profile. We have obtained the geometry and matter perturbations under linear homogeneous perturbation approach. These linear perturbations are observed to decay out gradually with the growth of the scale factor and ensure the stability of the model presented. Since the present Universe is accelerating, and the reason behind this is due to the presence of dark energy, we recreate the model as that of dark energy by incorporating a dimensionless parameter that includes the density parameter and  connect to the Hubble parameter. Subsequently the dimensionless parameter appears in the expression for the energy density and EoS parameter. As the EoS evolves from phantom phase to the quintessence phase, we obtained the specific value of the free parameters used in the dimensionless parameter for the $\Lambda$CDM model. We can conclude that the model confirms the bouncing behaviour and shows an accelerating behaviour in the dark energy era.  \\

As a final remark, we say that, we have presented a non-singular bouncing scenario within an extended gravity theory which is observed to be stable both in the pre and post bounce epochs. In general, a bouncing cosmology involves a contracting phase before bounce followed by a hot expanding phase. The contracting phase supposed to provide the mechanism to fatten and smooth the cosmological background and generate nearly scale-invariant density fluctuations spanning over length scale larger than the Hubble radius that act as seeds of structure in the post-bounce Universe. The period of slow contraction, also known as the ekpyrotic contraction phase, may be considered as a smoothing cosmological phase, where the Universe adheres to homogeneity, isotropy and  flatness.  However, in our model, we have not considered such smoothing issue. Within the formalism discussed in the present work, it still remains open to consider a period of slow contraction to address such issues.

\section*{Acknowledgement} ASA acknowledges the financial support provided by University Grants Commission (UGC) through Senior Research Fellowship (File No. 16-9 (June 2017)/2018 (NET/CSIR)), to carry out the research work. F. Tello-Ortiz thanks the financial by projects ANT--1956 and SEM 18--02 at the Universidad de Antofagasta, Chile. F. Tello-Ortiz acknowledges the PhD program Doctorado en Física mención en Física Matemática de la Universidad de Antofagasta for continuous support and encouragement. BM and SKT acknowledge IUCAA, Pune (India) for providing support through the visiting associateship program. The authors are thankful to the anonymous referee for the valuable comments/suggestions to improve the quality of the manuscript.\\

\end{document}